\documentclass{article}
\usepackage[nonatbib,final]{neurips_2019}

\usepackage[utf8]{inputenc} % allow utf-8 input
\usepackage[T1]{fontenc}    % use 8-bit T1 fonts
\usepackage{hyperref}       % hyperlinks
\usepackage{url}            % simple URL typesetting
\usepackage{booktabs}       % professional-quality tables
\usepackage{amsfonts}       % blackboard math symbols
\usepackage{nicefrac}       % compact symbols for 1/2, etc.
\usepackage{microtype}      % microtypography

\usepackage{graphicx}
\usepackage{xcolor}

\title{Dialog on a Canvas with a Machine}

\author{%
  Vivien Cabannes\thanks{Alphabetical order.}\\
  Ecole Normale Sup\'erieure \& INRIA\\
  Paris, France\\
  \And
  Tina Campana \footnotemark[1]\\
  \href{https://www.instagram.com/tinaetcharly/?hl=en}{\emph{Tina\&Charly}}\\
  Paris, France\\
  \And
  Charly Ferrandes \footnotemark[1]\\
  \href{https://www.instagram.com/tinaetcharly/?hl=en}{\emph{Tina\&Charly}}\\
  Paris, France\\
  \And
  Thomas Kerdreux\footnotemark[1] \\
  Ecole Normale Sup\'erieure \& INRIA\\
  Paris, France\\
  \And
  Louis Thiry \footnotemark[1] \\
  Ecole Normale Sup\'erieure\\
  Paris, France\\
}

\begin{document}

\maketitle

\begin{abstract}
We propose a new form of human-machine interaction. 
It is a pictorial game consisting of interactive rounds of creation between artists and a machine. 
They repetitively paint one after the other. 
At its rounds, the computer partially completes the drawing using machine learning algorithms, and projects its additions directly on the canvas, which the artists are free to insert or modify. Alongside fostering creativity, the process is designed to question the growing interaction between humans and machines.
\end{abstract}

\textbf{Creation Process.} 
With the on-going technological revolution, the human-machine interaction is deeply evolving. 
Hence art creation could benefit of new tools while simultaneously supporting thoughts of how these interactions are affecting humans.

Recently, GANs put a spotlight on the creative power of neural networks. 
For instance \cite{tan2017artgan,jin2017towards,donahuedisentangled,elgammal2017can,zhao2018emotiongan,chuartistic,lingenerating,grimm2018training,jetchev2018copy}
 were able to generate aesthetic full-stack painting. Yet in these, humans are either engineers or curators.
In this work, we propose a new utilisation of the machine, integrating it at the core of a human creative process.
The idea is to suggest to humans, while painting, ramifications and directions of their on-going artwork. 
In the following, we approach this generic idea under a specific interactive framework.

The artist duo \href{https://www.instagram.com/tinaetcharly/?hl=en}{\emph{Tina\&Charly}} have explored interaction using canvas as a media. 
To begin a creation, they choose a theme and symbolize it in dark on a white canvas. Then starts a game. 
At each round, using a basis of strokes and symbols that forms their pictorial vocabulary, % (Figure \ref{fig:comparaison_style}), 
Charly waits for Tina to schematize her emotions and thoughts in red, before answering her in green on the on-going painting. 
Rounds follow up until a consensus is reached about ending the painting. 
The whole process takes place in silence, the only dialog being on the canvas. 

The goal of this work is to introduce an artificial intelligence as a third player in \href{https://www.instagram.com/tinaetcharly/?hl=en}{\emph{Tina\&Charly}}'s dialog. 
The AI machine first captures a raw representation of the painting, then analyzes this signal to partially complete the on-going painting; completion that it projects back on the canvas. 
At this point, the artists are free to incorporate the machine's suggestion in blue, a color that has not been assigned to any player. 
At the end, having used different colors allows to analyze players’ contributions.

\begin{figure}
\centering
\includegraphics[width=.17\linewidth]{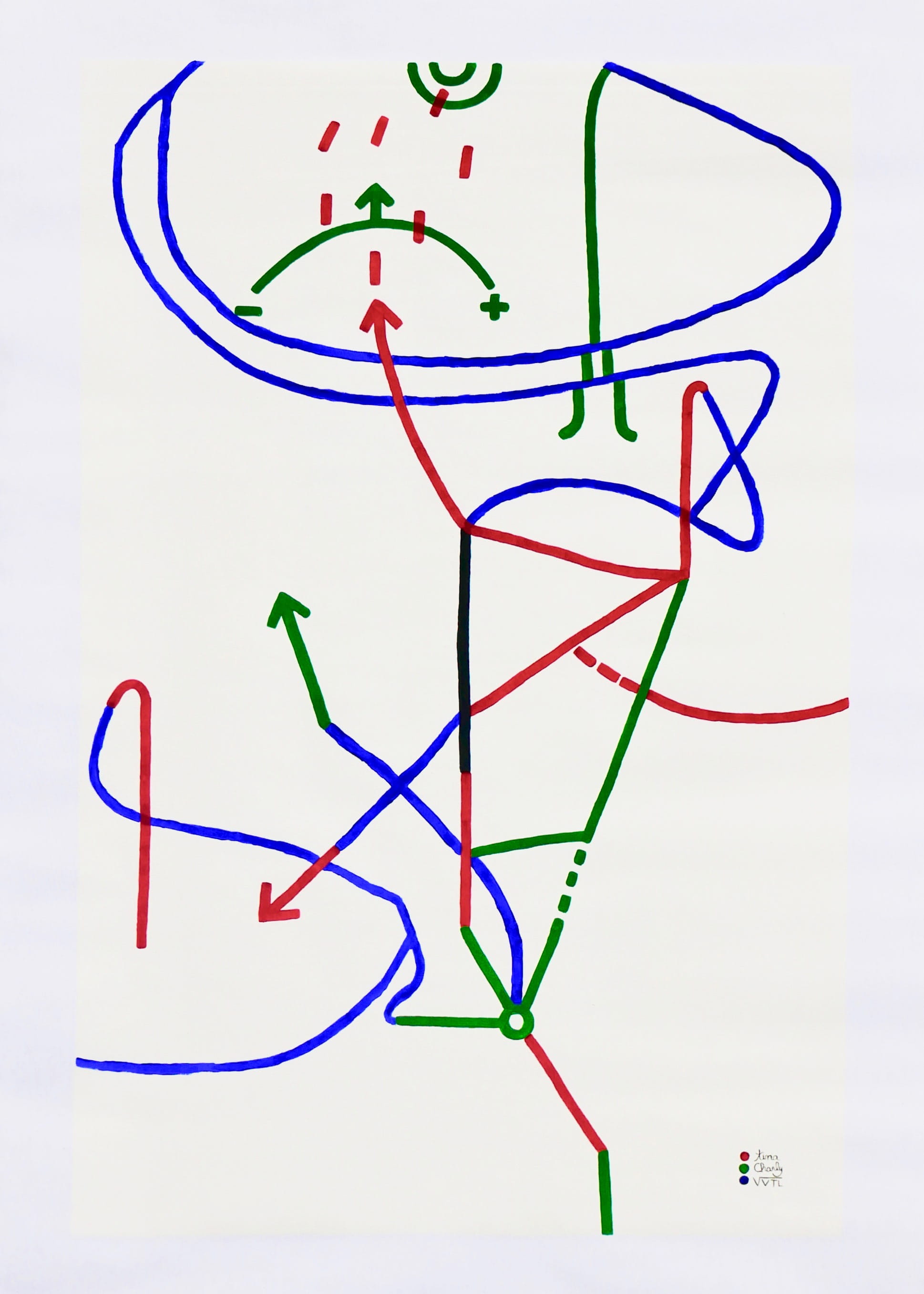}
\includegraphics[width=.17\linewidth]{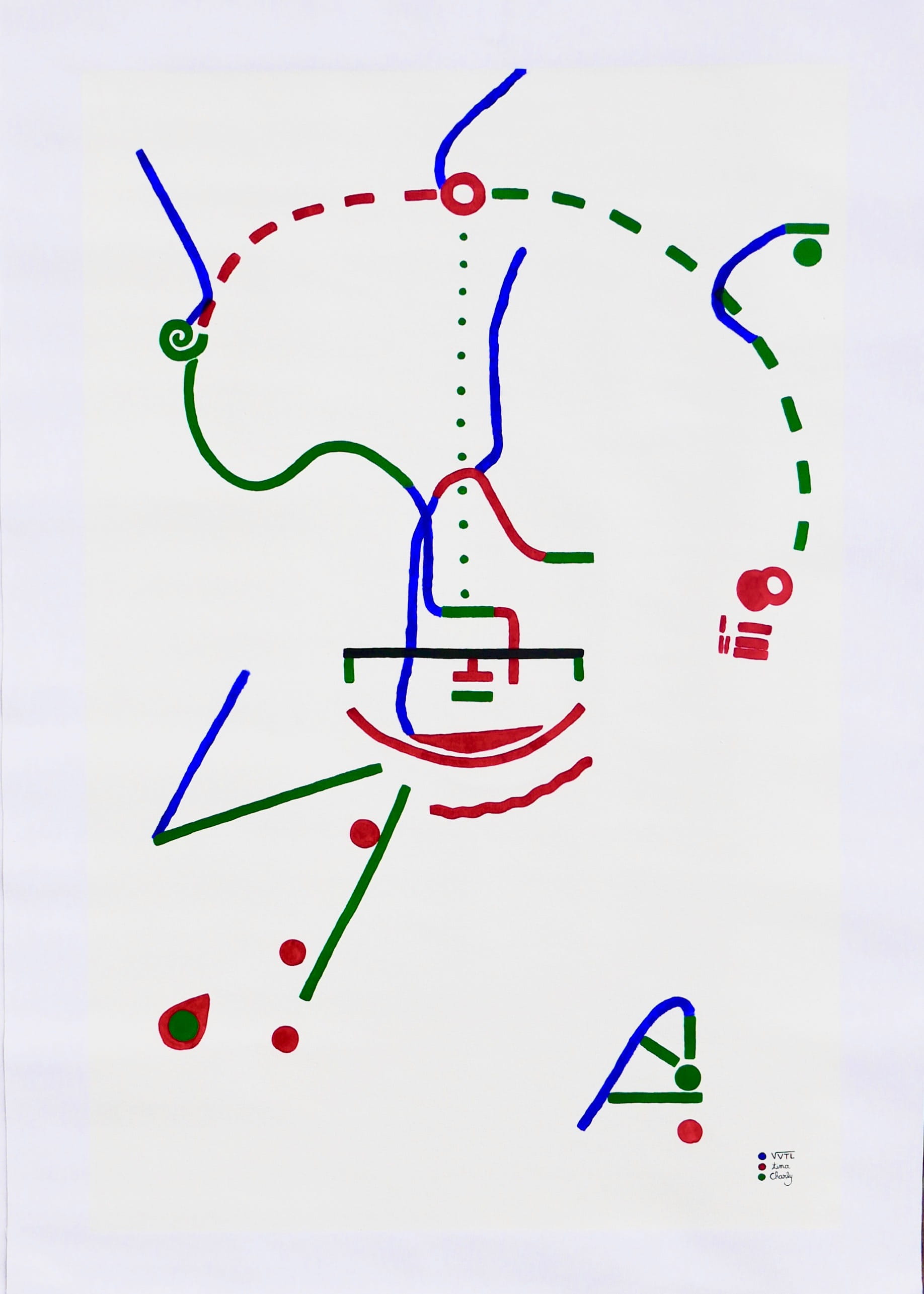}
\hspace{1cm}
\includegraphics[width=.17\linewidth]{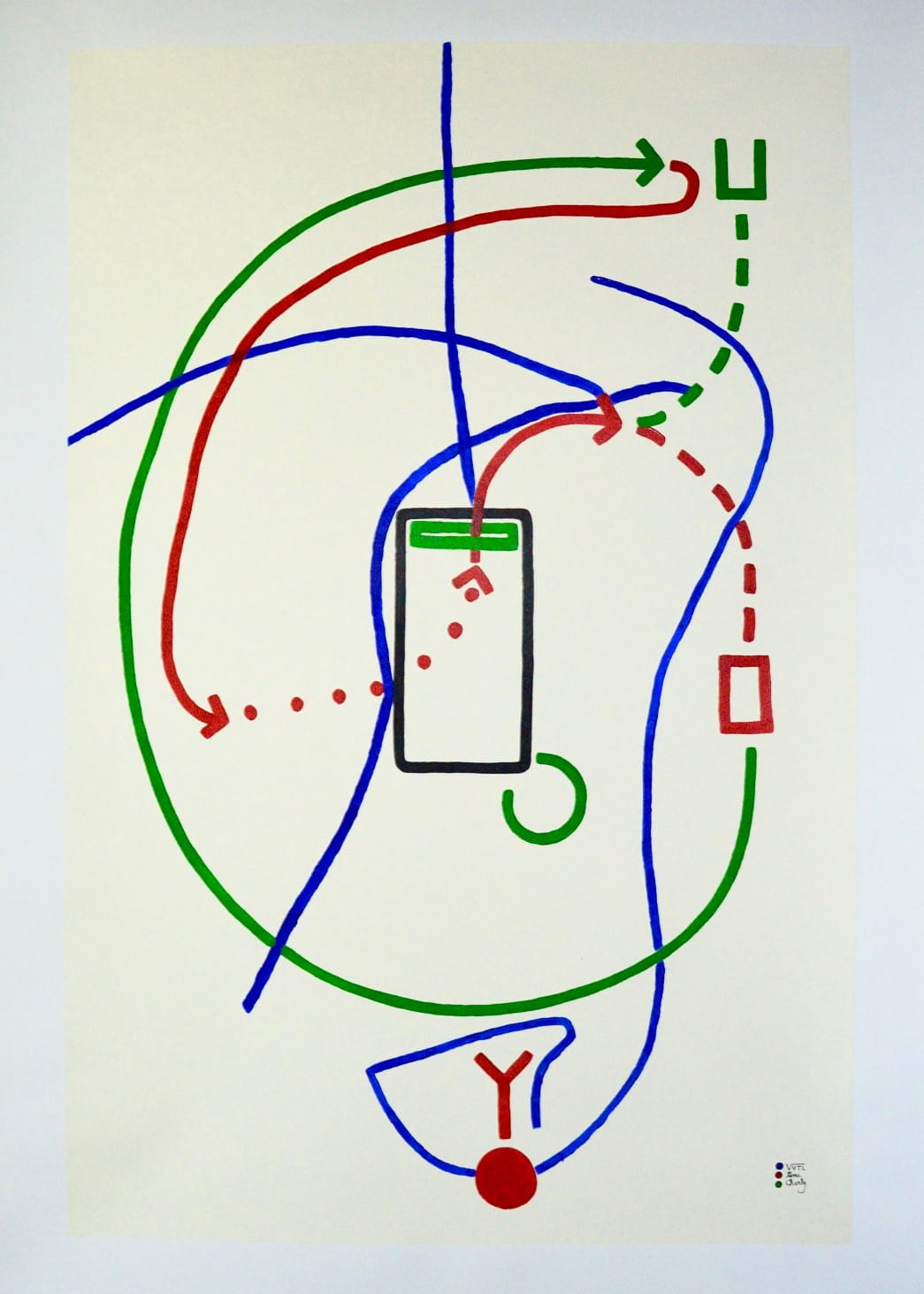}
\includegraphics[width=.17\linewidth]{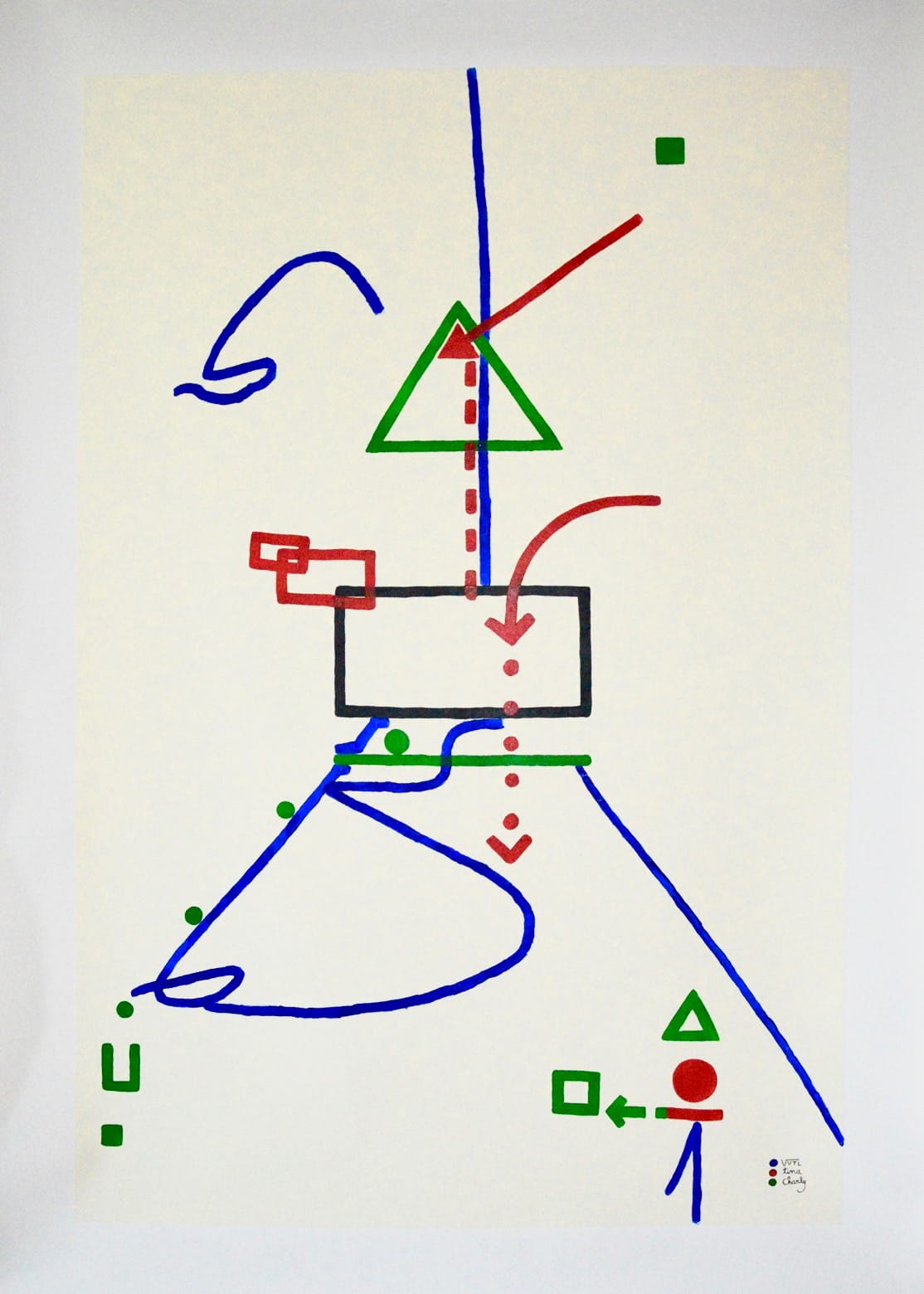}
\caption{Four 110 $\times$ 160 cm acrylic on canvas paintings. Two diptychs: \emph{Active, Passive} and \emph{Emitter, Receptor}, from left to right. The blue strokes are the computer projected suggestions interpreted by the artists. In \emph{Emitter} it only completes its own strokes or the black strokes of the canvas, while in its dual \emph{Receptor}, it completes any. It symbolizes a landmark of human-machine interaction, when human starts to systematically send back information that could condition the machine actions.}
\label{fig:actif_passif_emetteur_recepteur}
\end{figure}

\textbf{Installation and Specifications.}
The engineered system is composed of a camera and a projector connected to a computer on a still support (see Figure \ref{fig:installation}). 
At computer round, the system acquires an image of the painting and analyzes it to recover the exact canvas strokes.
This pre-processing was made robust to most luminosity variation for the interaction to be applicable in any studio in a seamless fashion.
Those strokes feed a \emph{neural sketcher}, that outputs new strokes to add on the painting. 
Finally post-processing allows to project those additions back on the canvas.

The neural sketcher is a recurrent neural network, based on recent powerful improvement \cite{ha2017neural}  of the seminal work of \cite{graves2013generating}. 
It is fed using doodling representation as a sequence of points along with a channel encoding for strokes breaks \cite{graves2013generating,ha2017neural}. 
The sketcher then outputs a similar series, that we convert back as strokes on the original painting. 
To train the network, we used the QuickDraw data set \cite{quickDraw}, it enables the network to produce human-like strokes. 
For a smoother integration with \href{https://www.instagram.com/tinaetcharly/?hl=en}{\emph{Tina\&Charly}} style, we further refined the learning using a sketch database from previous painting of the artists, collected by finding strokes of these and decomposing them into ordered points.

\textbf{Fostering Creativity.}
The artists found the machine strokes surprising and suggestive of move they would not have done by themselves.
Actually, some painters have expressed how evocative unintended strokes could be \cite[Chapter XII]{deleuze}.
Our installation where the machine projects completions without painting, combined with generative network capability, allows to explore that in a principled way.
Furthermore, the ability to change parameters, such as the learning data set or the amount of completion, adds more degree for the human to control their use of the machine.

\textbf{Human and Machine Interplay.}
Our physically interactive installation aims to be used by anybody, hoping to raise awareness and initiate thoughts on human and machine interplay. 
Arguably, it embodies that our use of technology is a middle ground where machines are made human-friendly and human drift from their original routines and spaces. 
Indeed, \href{https://www.instagram.com/tinaetcharly/?hl=en}{\emph{Tina\&Charly}} felt interacting with a full-body system -- it has been designed to superficially borrow as much as possible human-like painting behavior. 
They experienced the machine as sometimes constraining, hard to grasp, and sometimes magical, infusing new dimensions to the painting. 
Feeling, while in the creative process, that the machine could either be collaborative or muzzling, was an unexpected echo to what technologies seems to be in our daily life.

\begin{figure}
\centering
\includegraphics[width=0.17\linewidth]{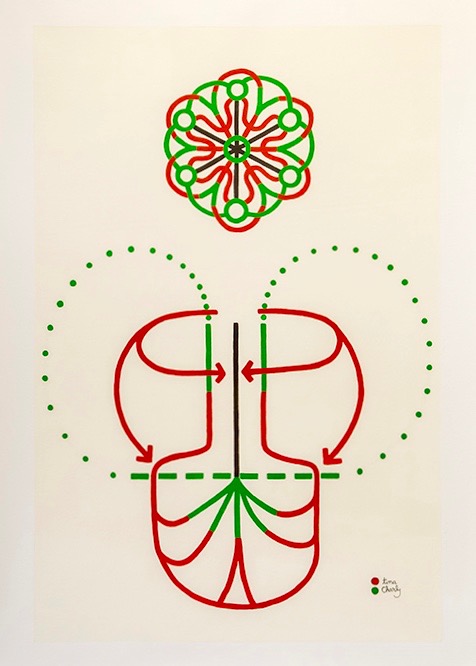}
\includegraphics[width=0.17\linewidth]{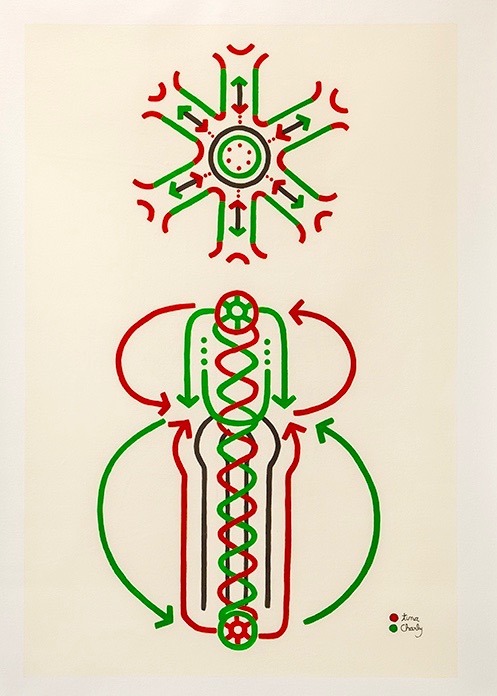}
\hspace{1cm}
\includegraphics[width=0.17\linewidth]{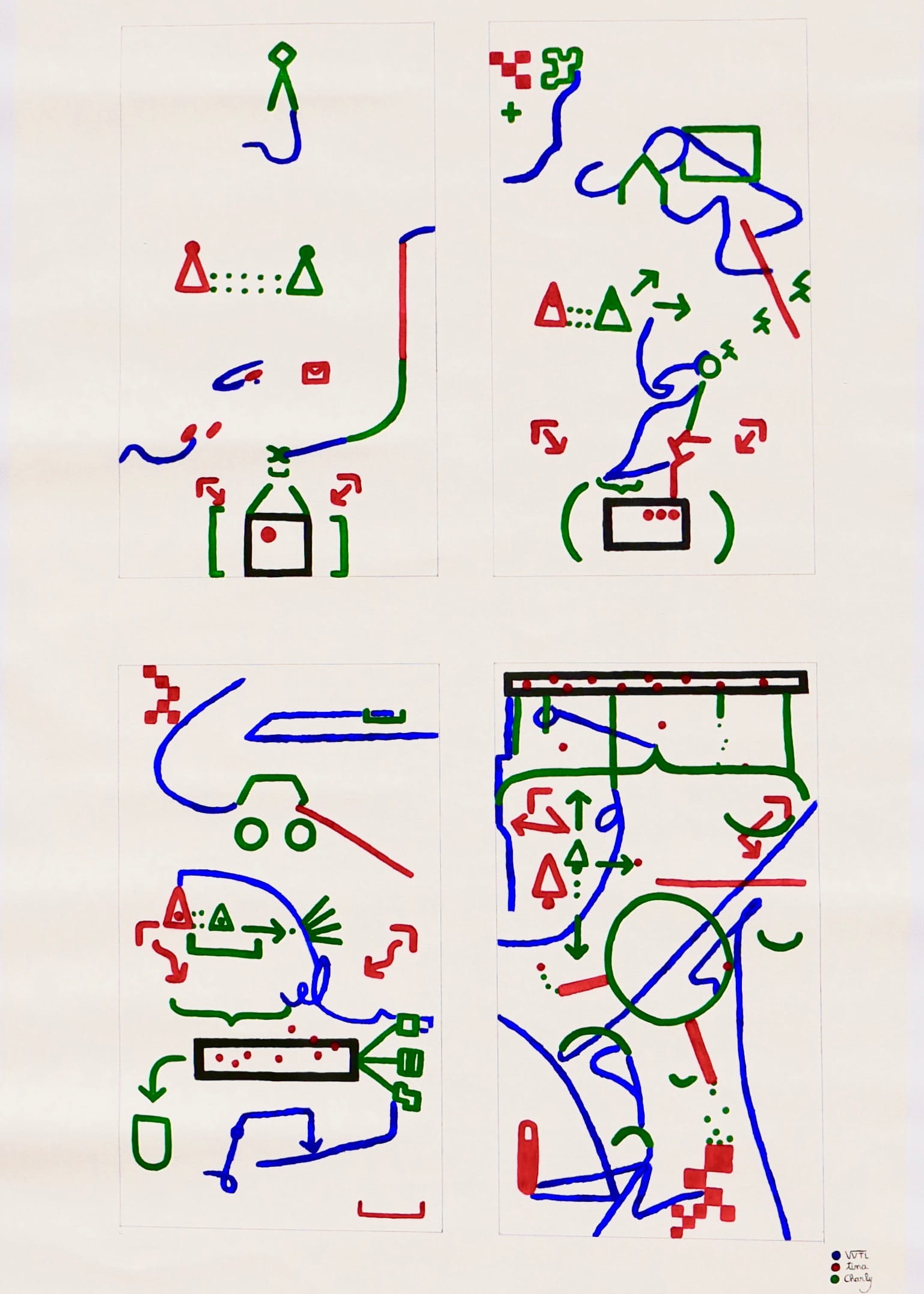}
\includegraphics[width=0.17\linewidth]{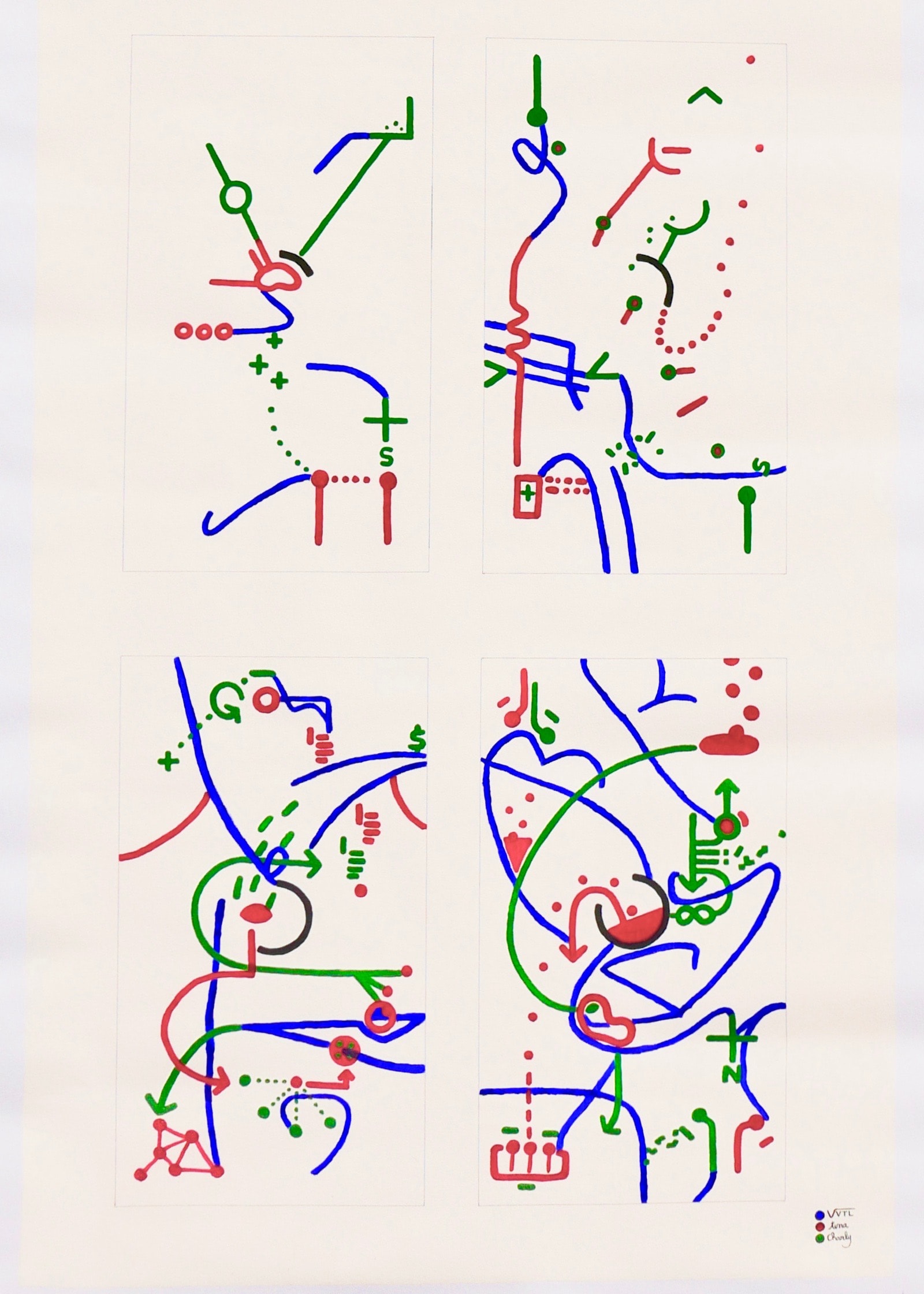}
\caption{Comparing the style of \href{https://www.instagram.com/tinaetcharly/?hl=en}{\emph{Tina\&Charly}} without, diptych on the left, or with the machine, diptych on the right. Acrylic on canvas, 110 $\times$ 160 cm.}
\label{fig:comparaison_style}
\end{figure}

From an outside perspective, the machine distorts their original painting style, both on the short term artworks resulting from their interaction (see Figure \ref{fig:comparaison_style}), and on their long term body of work as it inspired them on their machine-free paintings. 
As such, the interaction is not innocuous, even though, contrarily to our daily experience, we have made the machine impact as explicit as possible with its recognizable blue contributions.

\subsubsection*{Acknowledgments}
The authors would like to thanks Yana Hasson, Yann Labbé for some nice coding insights, Erwan Kerdreux for art history discussions and Thomas Lartigue for more general purposes discussions.

\bibliographystyle{plain}
\bibliography{biblio_art.bib}{}

\newpage
\appendix
\section{Additional images}

\begin{figure}[h!]
\centering
\includegraphics[width=.3\linewidth]{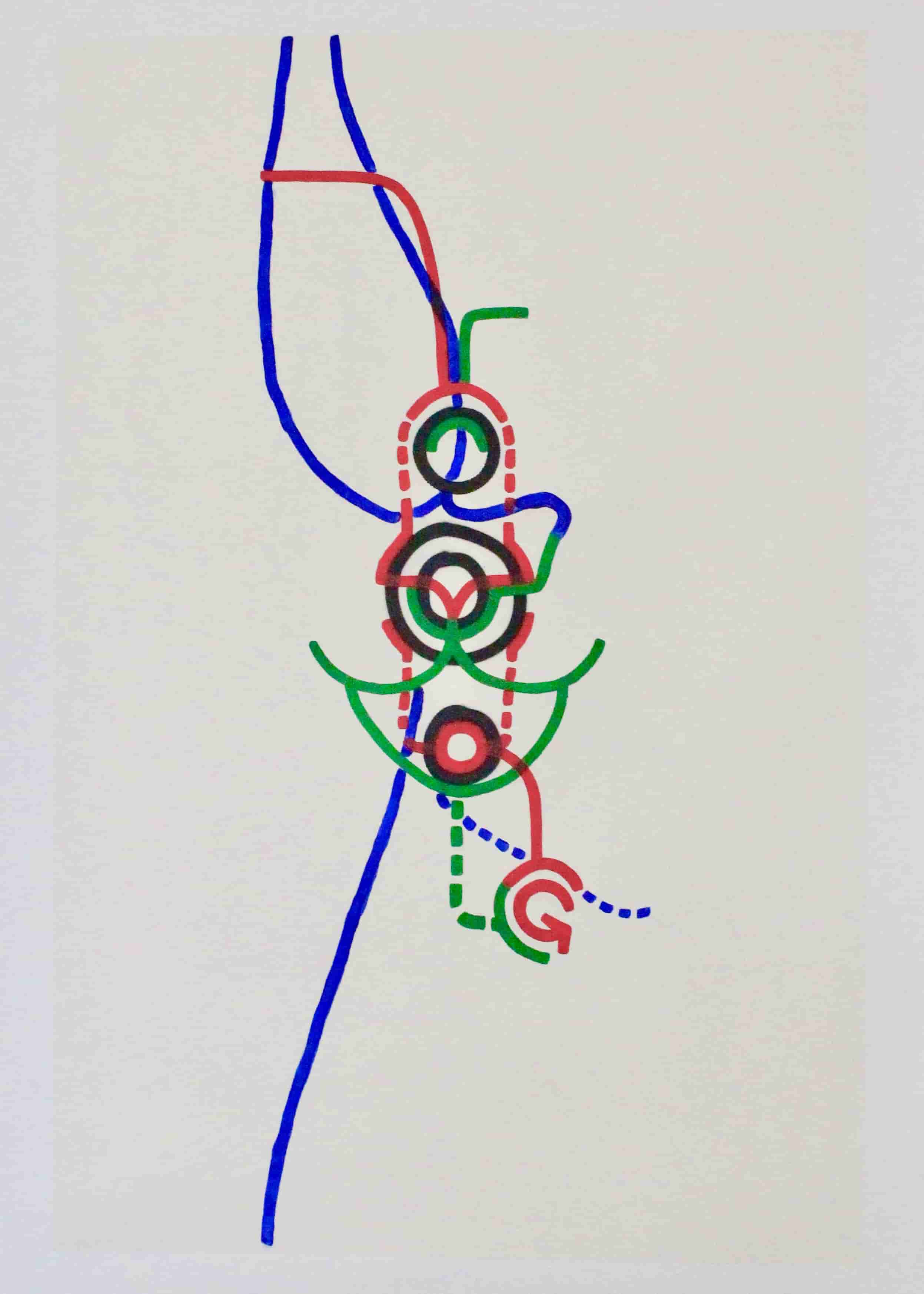}
\includegraphics[width=.3\linewidth]{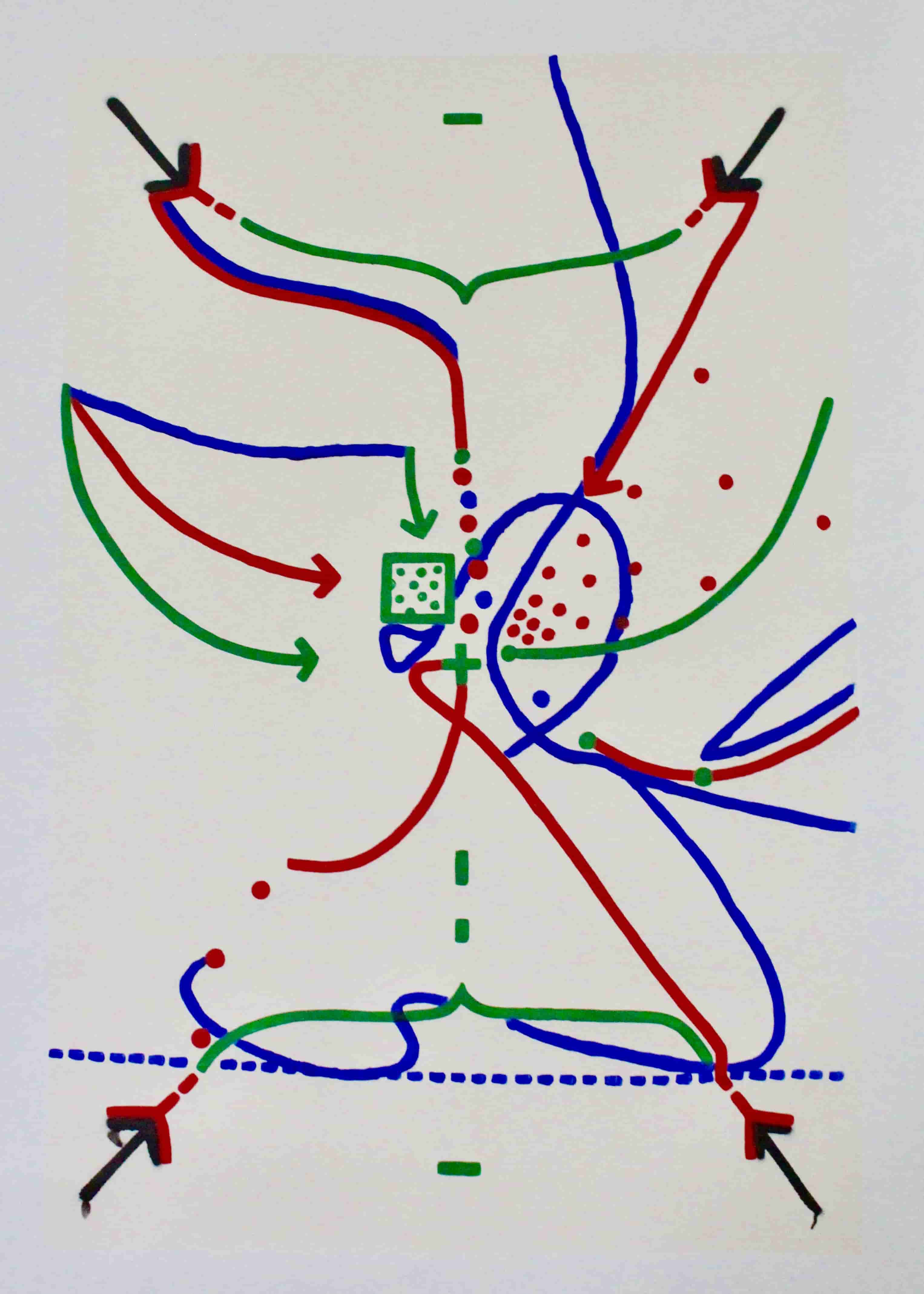}
\includegraphics[width=.3\linewidth]{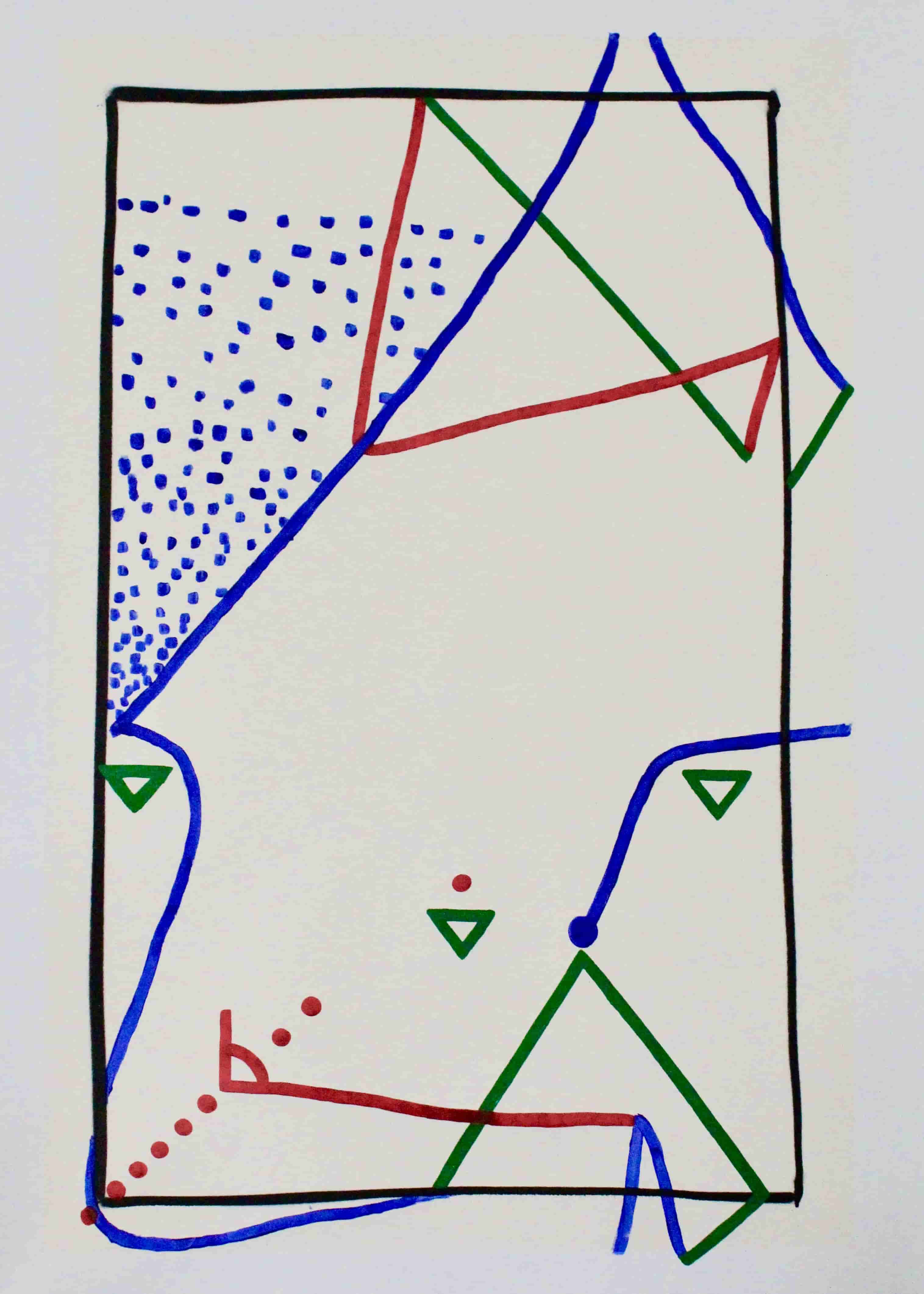}
\caption{Three acrylic on canvas paintings, each 110 $\times$ 160 cm. Those are the first three paintings of the series \emph{Influence, Convergence, Contr\^ole} and \emph{Monopole}.}
\label{fig:iccm}
\end{figure}

\begin{figure}[h!]
\centering
\includegraphics[width=.19\linewidth]{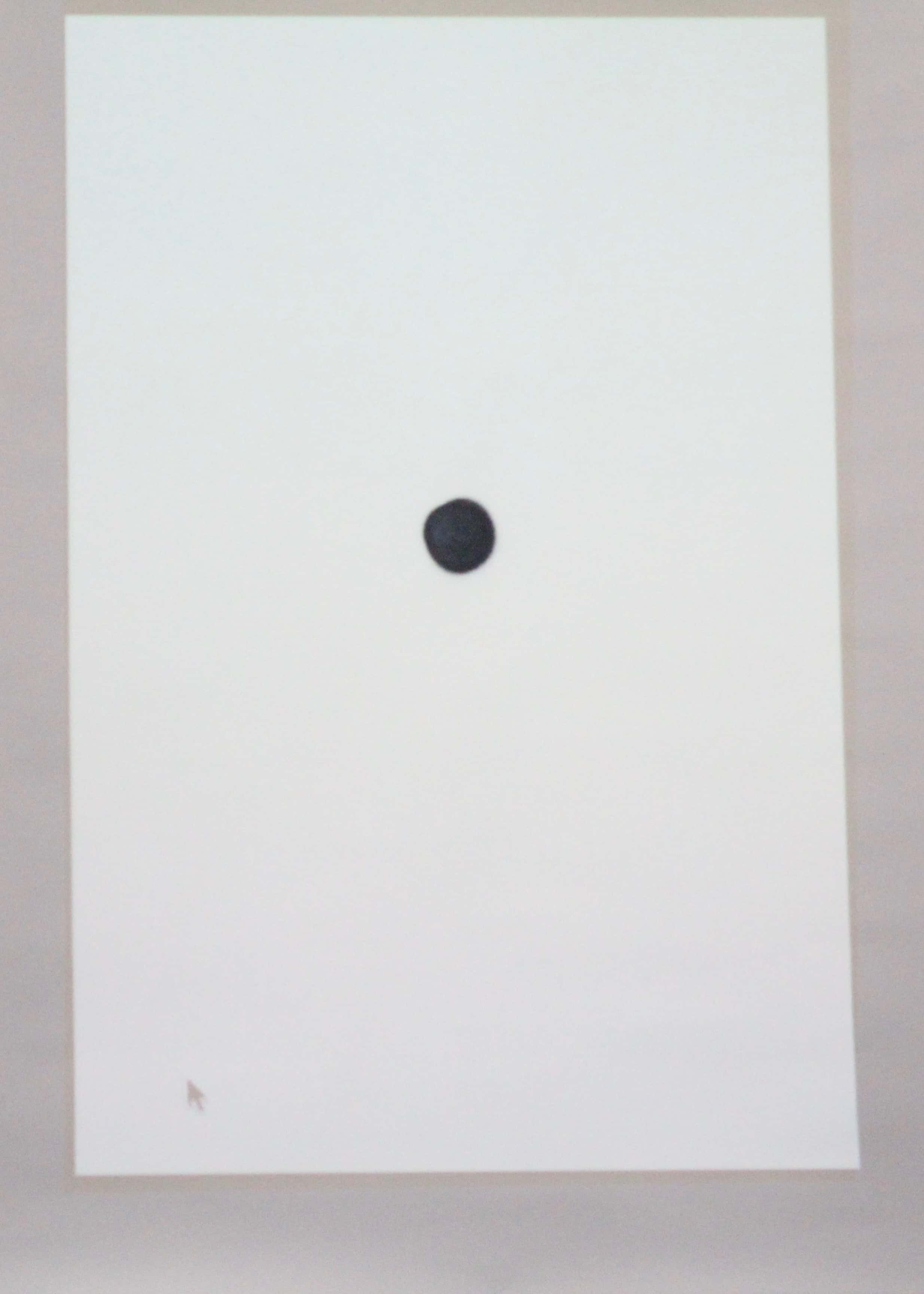}
\includegraphics[width=.19\linewidth]{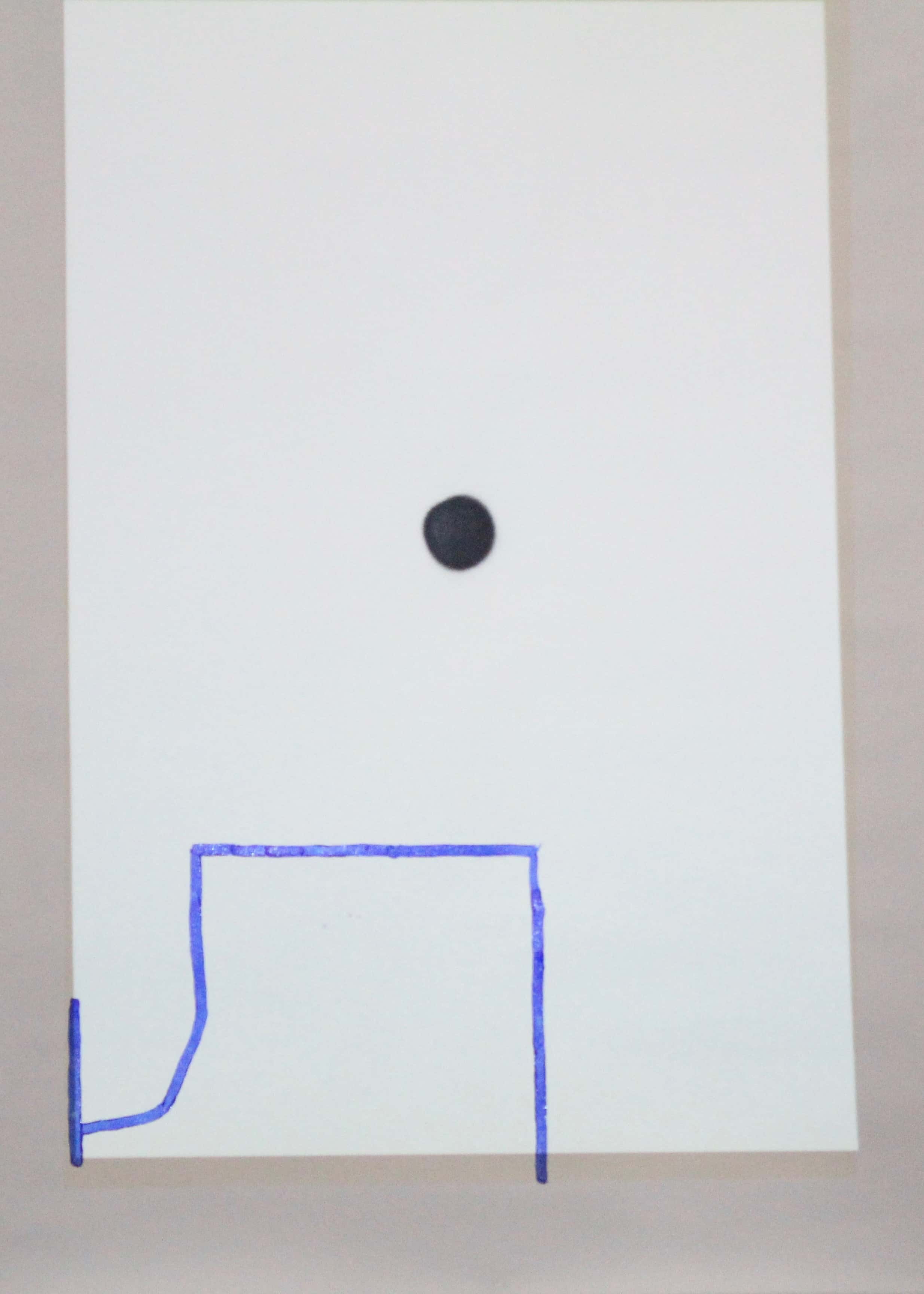}
\includegraphics[width=.19\linewidth]{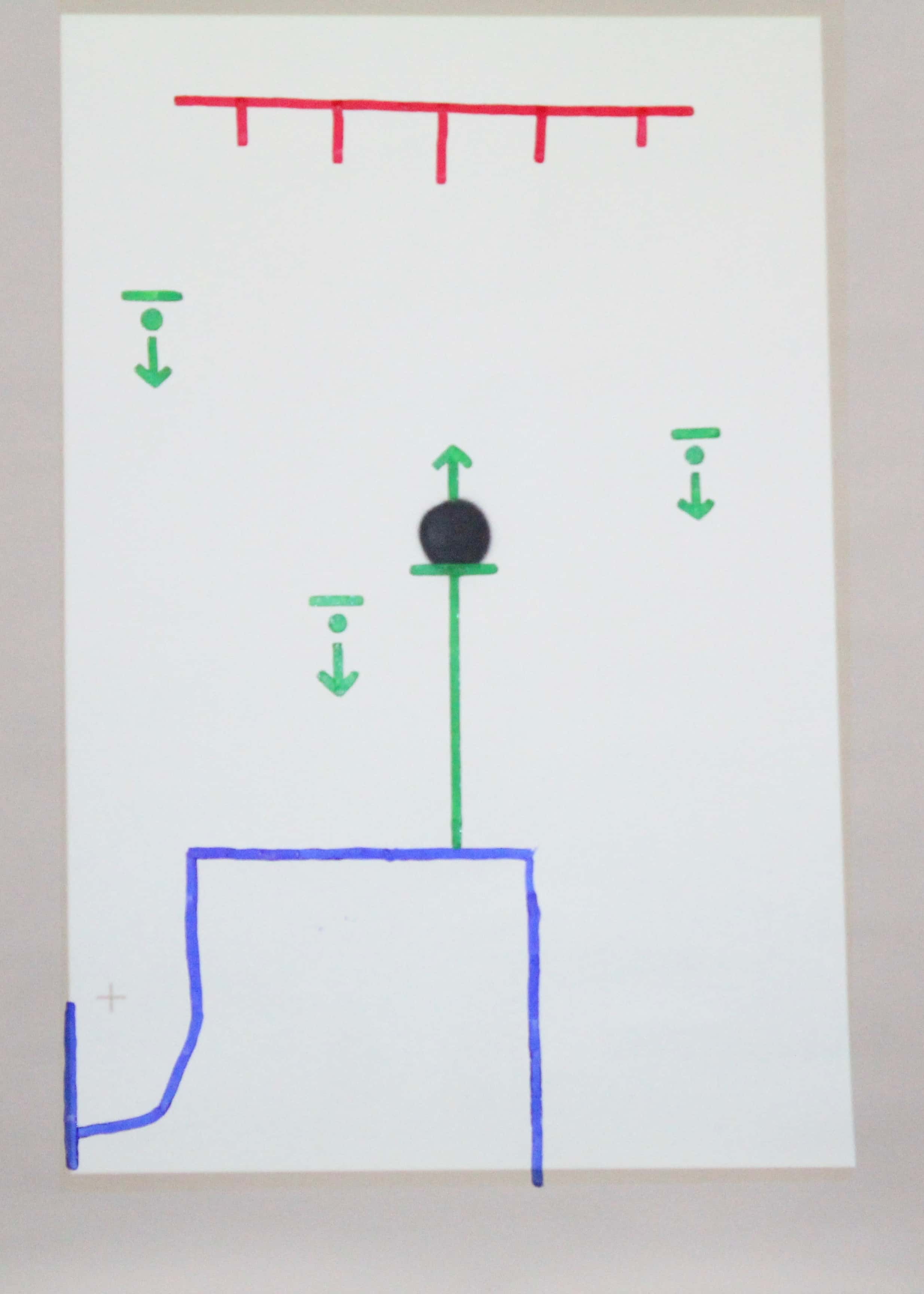}
\includegraphics[width=.19\linewidth]{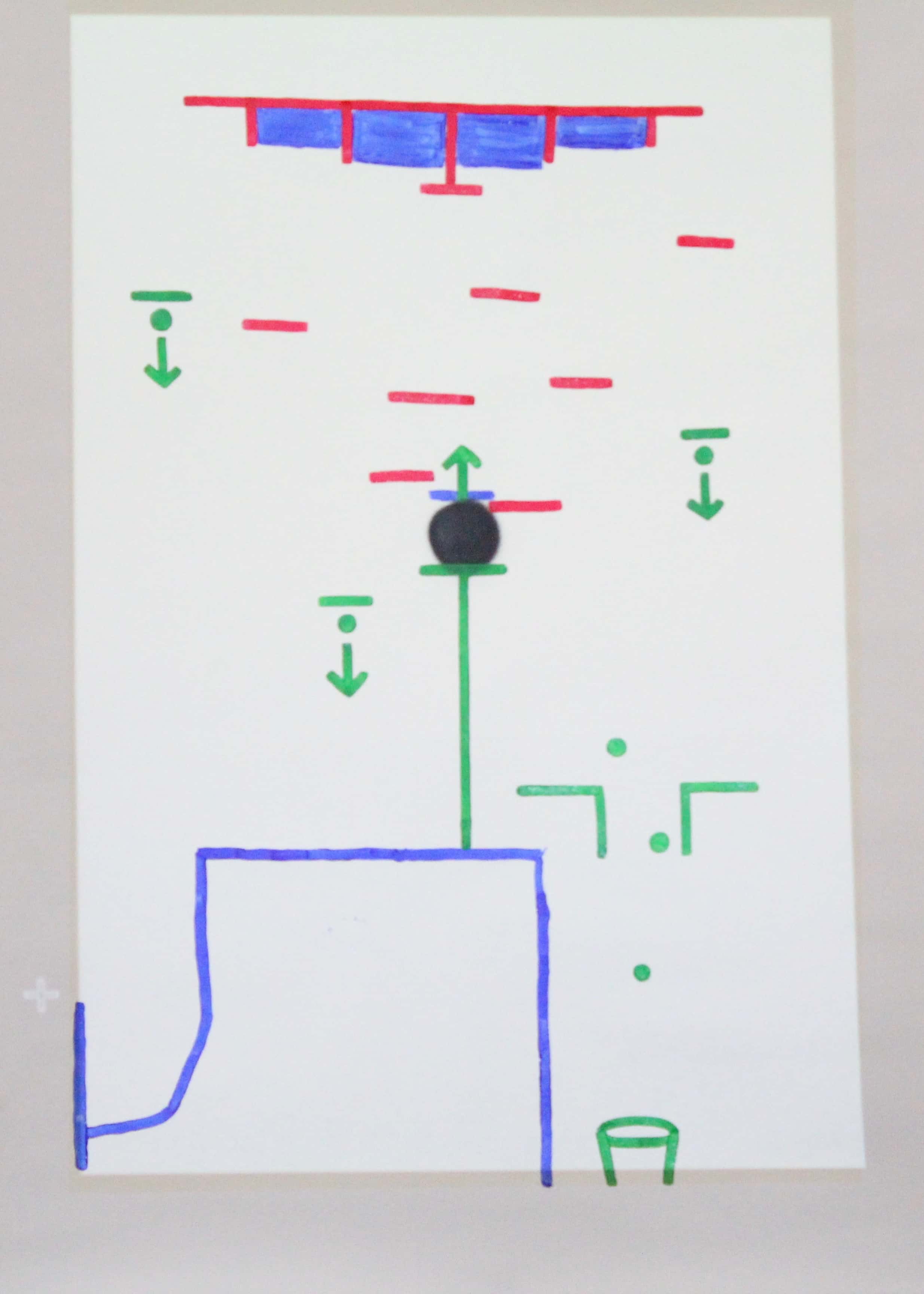}
\includegraphics[width=.19\linewidth]{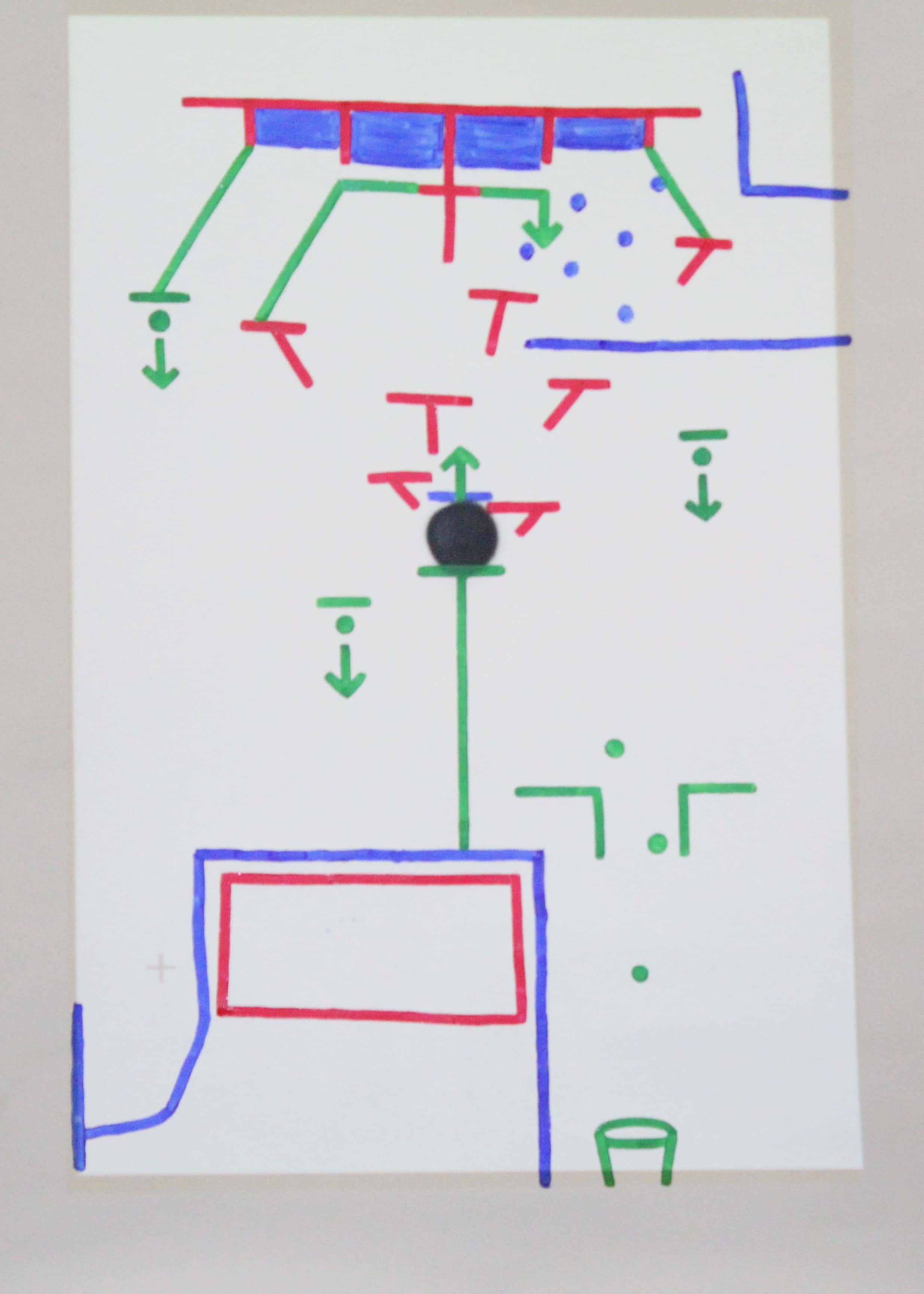}
\caption{Computer captures of the on-going \emph{Monopole} painting. Before taking the picture, the system projects white light on the canvas to have better lighting and ease the following processing steps.}
\label{fig:monopole}
\end{figure}

\begin{figure}[h!]
\centering
\includegraphics[width=.45\linewidth]{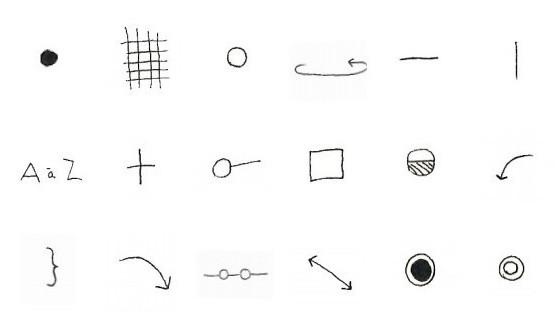}
\includegraphics[width=.45\linewidth]{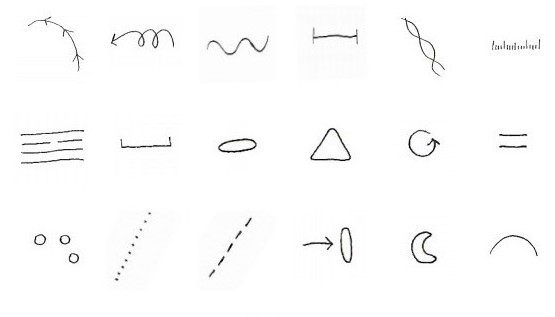}
\caption{\emph{Tina\&Charly} glossary, which could be given to a computer as a pictorial vocabulary.}
\label{fig:glossary}
\end{figure}

\begin{figure}[h!]
\centering
\includegraphics[width=.32\linewidth]{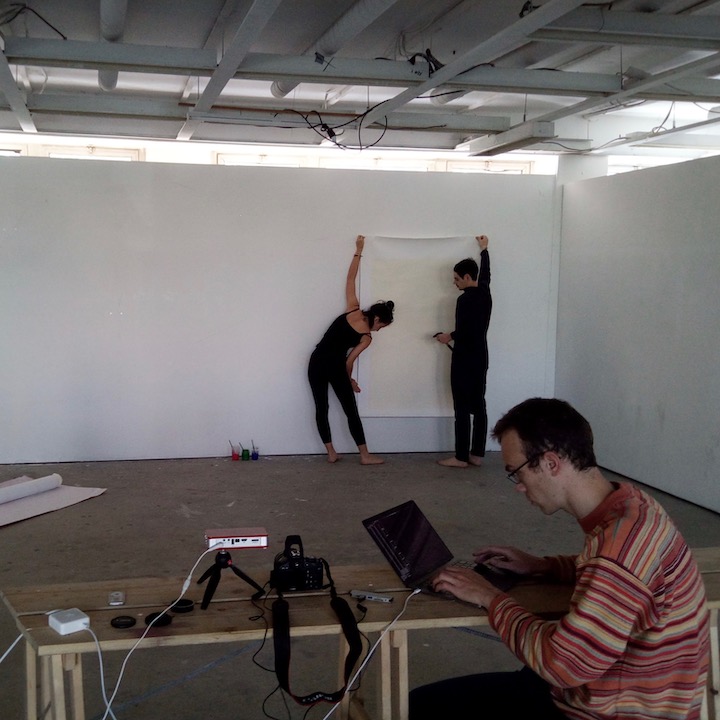}
\includegraphics[width=.32\linewidth]{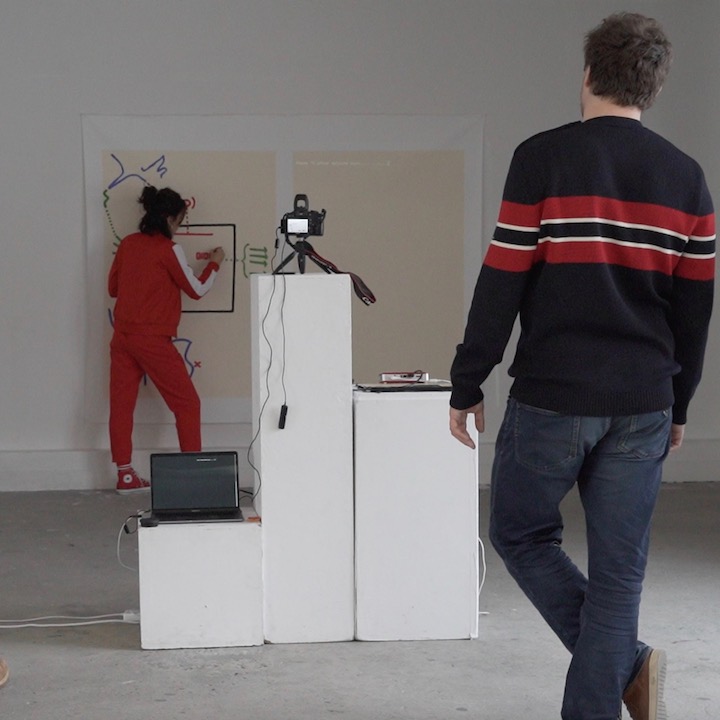}
\includegraphics[width=.32\linewidth]{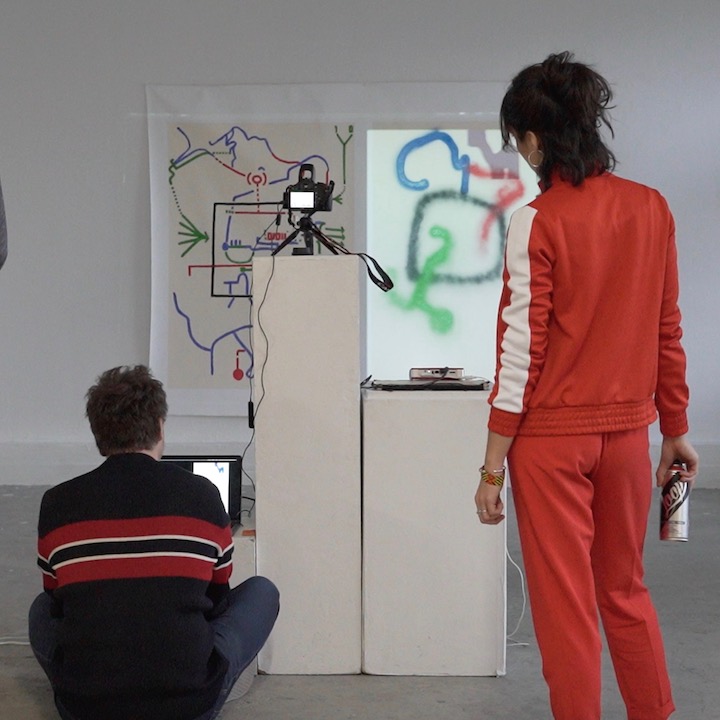}
\includegraphics[width=.32\linewidth]{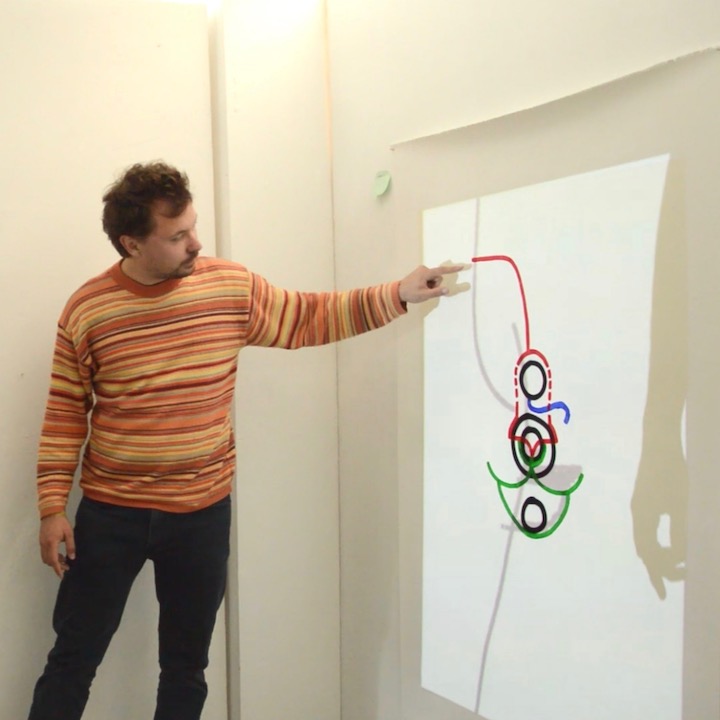}
\includegraphics[width=.32\linewidth]{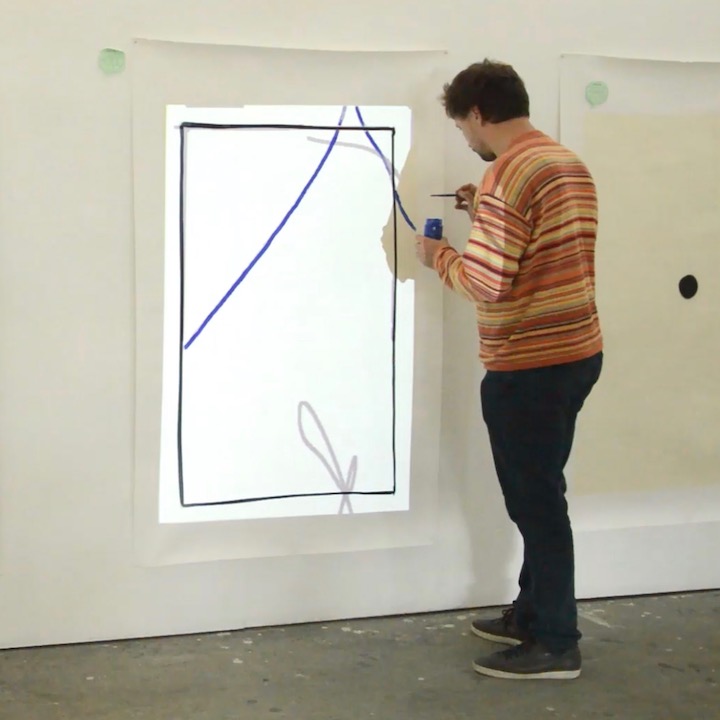}
\includegraphics[width=.32\linewidth]{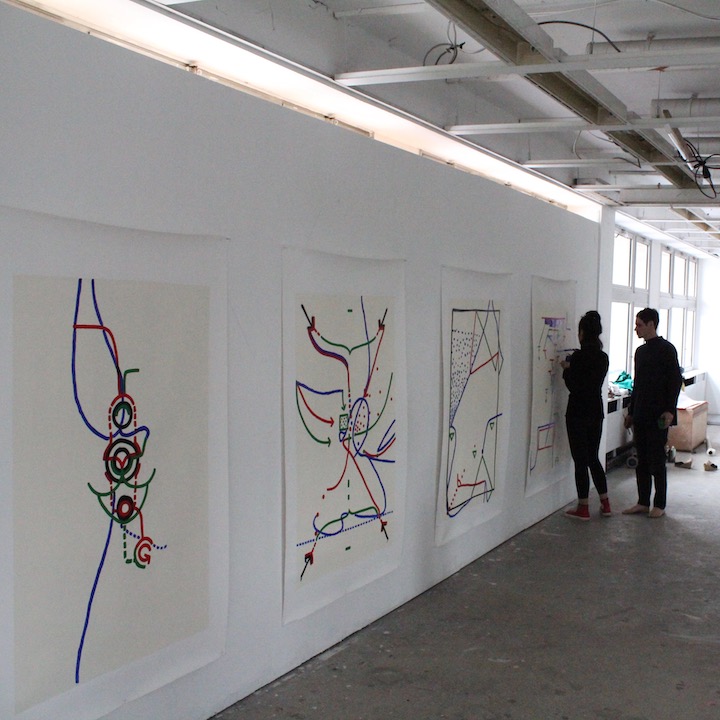}
\caption{Some images of the painting process in Atelier 6B, Saint-Denis, France. (Top left) Artists install canvas while computer scientists install their machine. The machine is made of a camera, a computer and a projector, it is highly portable. (Top middle) The artist draws under the scrutiny of the computer. (Top left) The computer analyzes the on-going painting in order to suggest additions. (Bottom left) Those suggestions are projected on the canvas for the artists to discuss addition. (Bottom middle) Additions are incorporated in blue on the canvas. (Bottom right) At the end, the artists apply a glaze mixture to protect their creations.}
\label{fig:installation}
\end{figure}

\end{document}